# Fluctuations in DNA Packing Density Drive the Spatial Segregation between Euchromatin and Heterochromatin

Luming Meng[2*], Boping Liu[2], and Qiong Luo[1*]

[1]Center for Computational Quantum Chemistry, School of Chemistry, South China Normal University, Guangzhou 510631, People's Republic of China

[2]Materials and Energy of Ministry of Education, College of Materials and Energy, South China Agricultural University, Guangzhou 510630, People's Republic of China

* Luming Meng and Qiong Luo

**Email:** menglum@scau.edu.cn (L.M.), luoqiong@scnu.edu.cn (Q. L.)

**Abstract**

In the crowded eukaryotic nucleus, euchromatin and heterochromatin segregate into distinct compartments, a phenomenon often attributed to homotypic interactions mediated by liquid–liquid phase separation (LLPS) of chromatin-associated proteins. Here, we revisit genome compartmentalization by examining the role of in vivo DNA-packing density fluctuations driven by ATP-dependent chromatin remodelers. Leveraging DNA accessibility data, we develop a polymer-based model that captures these fluctuations and successfully reproduces genome-wide compartment patterns observed in Hi-C data, without invoking homotypic interactions. Further analysis reveals that density fluctuations in a crowded nuclear environment elevate the system's energy, while euchromatin–heterochromatin segregation facilitates energy dissipation, offering a thermodynamic advantage for spontaneous compartment formation. These findings suggest that euchromatin–heterochromatin segregation may arise through a non-equilibrium, self-organizing process, providing new insights into genome organization.

**Introduction**

The eukaryotic nucleus is a highly crowded and spatially constrained environment, exemplified by the compaction of nearly two meters of human genomic DNA into a nucleus just 6–10 μm in diameter(*1*). During interphase, this DNA is packaged into chromatin, a non-uniform linear array of nucleosomes, each consisting of ~147 base pairs of DNA wrapped around a histone octamer(*1-6*). Chromatin exists in two distinct forms: loosely packed euchromatin, which exhibits dynamic, low DNA-packing densities that facilitate DNA accessibility and transcription, and heterochromatin, which is more densely packed, maintaining a relatively stable, high DNA-packing density and is typically associated with gene repression(*7-9*). A hallmark of nuclear organization is the spatial segregation of these two forms into active and inactive compartments(*10-12*). Yet, the physical mechanisms underlying this segregation remain unresolved(*13*).

Inspired by the observation that many chromatin-binding proteins can self-associate and form condensates via liquid-liquid phase separation (LLPS) *in vitro(14-16)*, a surge of research has examined genome compartmentalization through this lens (*11, 17*). It is now widely recognized that LLPS-driven self-association of these proteins promotes the spatial clustering of their target chromatin regions, providing a plausible physical mechanism for genome compartmentalization(*18-20*). However, it remains unclear whether current *in vitro* evidence is sufficient to demonstrate that a specific protein indeed undergoes LLPS to form condensates *in vivo*, making it premature to conclude that genome compartmentalization is driven by this mechanism under physiological conditions(*13, 21*).

Here, we turn to *in vivo* evidence implicating ATP-dependent chromatin remodelers as key regulators of chromatin dynamics(*1*). These remodelers actively reposition nucleosomes within euchromatin—by sliding, inserting, or evicting them—resulting in dynamic fluctuations in local DNA-packing density and sharpening its contrast with the more stably packed heterochromatin(*22, 23*). Live-cell imaging studies further reveal that such remodeling activities enhance large-scale chromatin mobility(*24*). These observations raise an important question: Can remodeler-driven modulation of nucleosome positioning actively shape 3D genome architecture? More specifically, do fluctuations in DNA-packing density mediated by chromatin remodelers contribute to the segregation of euchromatin and heterochromatin?

Addressing this question is challenging, as direct genome-wide measurements of DNA-packing density fluctuations are currently limited. Nevertheless, chromatin accessibility data obtained from ATAC-seq and DNase-seq experiments(*25, 26*) provide informative proxies for local DNA-packing density(*27-30*). In our previous work, we developed a polymer-based model to quantify local DNA-packing density from accessibility data, representing chromosomes as heteropolymers with a static distribution of DNA-packing density(*31*). While this model successfully captured the heterogeneous landscape of chromatin compaction, it did not account for dynamic fluctuations in DNA-packing density driven by chromatin remodelers. Here, we extend this framework to incorporate remodeling-induced fluctuations in DNA-packing density. Our results reveal that such fluctuations can drive the spontaneous segregation of euchromatin and heterochromatin within the crowded, spatially constrained nuclear environment.

**A polymer model to investigate the effect of DNA-packing density fluctuations on 3D genome organization**

In the crowded eukaryotic nucleus, ATP-dependent chromatin remodelers dynamically reposition nucleosomes within euchromatin, generating fluctuations in local DNA-packing density (*22,23*).

Here, we develop a polymer model that leverages DNA accessibility data to represent chromosomes as polymers, explicitly capturing these fluctuations without assuming attractions between chromatin regions.

Briefly, we develop the model in three steps (Fig. 1 and Methods). First, we represent the chromatin region of interest as a self-avoiding polymer chain by dividing it into consecutive segments of equal genomic length. Each segment is modeled as a series of contiguous, coarse-grained, self-avoiding beads of equal physical volume (Fig. 1A). Each bead is assigned two parameters: (1) the genomic sequence length it represents, and (2) the accessibility signal value ($\zeta$), which corresponds to the sum of accessibility reads mapped to that sequence. For clarity in the following description, beads within each segment are distinguished by color: the first bead is orange, while the others are cyan (Fig. 1A). For simplicity, our model initially represents each segment with a single bead, which also corresponds to the first bead in the segment and is marked in orange (Methods).

Second, we simulate the random folding of the polymer in a confined space and periodically perturb the number of beads in each segment during the simulation. At each perturbation step, we first calculate the accessibility signal ($\zeta$) for each bead and then determine its fate based on its $\zeta$ value. For the first bead of each segment colored in orange, it has two possible fates in the perturbation step: either remain unchanged or split evenly into an orange and a cyan bead, with the new orange bead always preceding the new cyan bead to ensure that the first bead in each segment remains orange (Fig. 1A). Notably, "split evenly" means that a bead divides into two new beads, each having the same physical volume as the original bead but representing half of its genomic sequence. For the other beads in each segment colored in cyan, they have three possible fates in the perturbation step: they can (i) disappear, (ii) remain unchanged, or (iii) split evenly into two new cyan beads, with probabilities dictated by their $\zeta$ values. If a cyan bead disappears, its genomic sequence is reassigned to the preceding bead (orange or cyan) to maintain continuity. As a result of these periodic perturbations, both the number of beads per segment and the values of the two parameters for each bead undergo dynamic changes during the folding simulation. As the number of beads in a segment increases, the segment occupies a larger spatial volume, resulting in a reduced DNA-packing density. Conversely, a decrease in bead number leads to higher DNA-packing density. These periodic perturbations in bead number simulate fluctuations in local DNA-packing density.

Third, the folding simulation involving periodic perturbations in bead number is repeated multiple times, each starting from a distinct initial conformation, to generate independent folding trajectories. Conformations are sampled from these trajectories to create an ensemble, which is then used to construct an ensemble-averaged contact matrix for comparison with Hi-C data (Fig. 1B, 1C, and Methods)

**Prediction of genome compartmentalization at a genome-wide scale**

We investigated the role of DNA-packing density fluctuations in 3D genome organization using the K562 human erythroleukemia cell line. Each of the 22 autosomes and the X chromosome was divided into consecutive 50-kb segments (Methods). Bulk ATAC-seq data from K562 cells(*32*) were used to generate chromatin conformation ensembles and their corresponding ensemble-averaged contact matrices for each chromosome. Simulated ensemble-averaged contact matrices displayed characteristic plaid patterns, indicative of A/B compartmentalization (Fig. 2B). To sharpen the plaid patterns, correlation matrices were constructed from the contact data (Fig. 2C). Principal component analysis (PCA) was subsequently applied to identify A and B compartments(*10*), generating compartment PC1 values (Fig. 2C and Figure S3). The same analysis was performed on experimental Hi-C data(*33*) for comparison (Fig. 2B, 2C, and Figure S3). The strong agreement between simulated and Hi-C contact matrices is supported by high Pearson correlation coefficients between their compartment PC1 values: 0.860 on chromosome 20 and 0.818 genome-wide (Fig. 2D and 2E). Furthermore, the simulated 3D conformations showed preferential spatial clustering of genomic regions belonging to the same compartment type (Fig. 2A), confirming that the simulated plaid patterns correspond to enriched homotypic-compartment contacts. These findings suggest a strong link between DNA-packing density fluctuations and chromosome compartmentalization.

To test the robustness of this link, we applied our model to predict chromatin contacts during early mammalian development. Using chromatin accessibility data from Du et al.'s study on preimplantation mouse development(*34*), we generated chromatin conformation ensembles for all 19 autosomes at both the 8-cell and inner cell mass (ICM) stages, along with their corresponding ensemble-averaged contact matrices and correlation matrices (Methods). Comparisons with experimental Hi-C data(*35*) revealed a strong correlation, further supporting the link between DNA-packing density fluctuations and chromosome compartmentalization. Specifically, for chromosome 19, the Pearson correlation coefficients between simulated and Hi-C compartment PC1 values were 0.724 at the 8-cell stage and 0.831 at the ICM stage (Figure S1D and S2D). Similarly, genome-wide compartment PC1 profiles exhibited correlations of 0.749 and 0.799 at the respective stages (Figure S1E and S2E).

Together, our simulations, based on the model considering only the DNA-packing density fluctuations within euchromatin induced by ATP-dependent chromatin remodelers, reveal that these fluctuations drive the spontaneous spatial segregation of euchromatin and heterochromatin during random folding, without requiring homotypic interactions.

**DNA-packing density fluctuations within euchromatin drive the spatial segregation of euchromatin and heterochromatin**

To investigate the underlying mechanisms of this spontaneous spatial segregation, we designed a 50-Mb artificial chromatin fiber composed of 1,000 50-kb segments, organized into 10 alternating red and blue blocks, each spanning 100 segments (Fig. 3A and Methods). Segments within red blocks are modeled using one or two beads, while segments in blue blocks are represented by a single bead. As a result, red blocks maintain a lower DNA-packing density, resembling euchromatin, compared to blue blocks, which mimic heterochromatin. This alternating red-blue block design reflects the distribution of euchromatin and heterochromatin in chromosomes.

Based on the 50-Mb artificial chromatin fiber, we constructed a dynamic model in which the DNA-packing density of red euchromatin-like blocks fluctuates, while the density of blue heterochromatin-like blocks remains constant throughout the folding simulation (Fig. 3A). Specifically, half of the red segments were randomly selected and assigned two beads each, while the remaining red segments and all blue segments were assigned one bead, resulting in a polymer model consisting of 1,250 beads for the artificial fiber. The polymer was then randomly folded in a confined space under periodic perturbations in the bead number of red segments. At each perturbation step, half of the red segments were re-randomly selected and assigned two beads each, while the remaining red segments received one bead, ensuring that the total bead count remained fixed. This process allowed red segments to alternate between one and two beads multiple times, generating dynamic fluctuations in the DNA-packing density of red blocks during the folding simulation, while the blue blocks remained unchanged. As a control, we also generated a static model by randomly folding the initial polymer chain consisting of 1,250 beads, without applying bead-number perturbations to the red segments during the folding simulation (Fig. 3A and Methods).

For both the dynamic and static models, we performed multiple folding simulations with distinct initial conformations, generating independent folding trajectories. From these trajectories, we sampled conformations to quantify the degree of intermingling between red and blue blocks (Fig. 3B and Methods). We found that in the static model, the red and blue blocks remained interspersed throughout the simulation, whereas in the dynamic model, they rapidly separated (Fig. 3B and Movies 1-5). Consistently, the ensemble generated from the sampled conformations revealed no characteristic plaid patterns indicative of segregation in its ensemble-averaged contact and correlation contact matrices of the static model, while distinct plaid patterns were observed in the matrices of the dynamic model (Fig. 3C). These results suggest that static differences in DNA-packing density between red euchromatin-like and blue heterochromatin-like blocks are insufficient to drive their spatial segregation. Instead, perturbations in bead number within red segments, which induce fluctuations in the DNA-packing density of red euchromatin-like blocks, promote their segregation from blue blocks.

To examine how fluctuations in the DNA-packing density of red blocks drive spontaneous red-blue block segregation, we analyzed the energy landscape of folding trajectories from dynamic model simulations. Each perturbation increased the system's energy, driving it out of equilibrium. This excess energy was subsequently dissipated through random folding before the next perturbation, allowing the system to relax toward a lower-energy conformation. Consequently, each trajectory consisted of repeated perturbation-relaxation cycles. By comparing the average energy dissipation curves between the early and late stages of the simulations (Fig. 3D), we found that the system's energy declined more rapidly in the late stage, when red and blue blocks were fully segregated into distinct domains, than in the early stage, when only partial segregation was observed. This observation suggests that segregated conformations promote more efficient energy dissipation following fluctuations in DNA-packing density, providing an energetic advantage that drives their spontaneous formation during random folding.

**Discussion**

An intriguing question emerging from our findings is why segregated conformations enhance energy dissipation following fluctuations in DNA-packing density. To investigate this, we first examined how perturbations in bead number within red blocks elevate the system's energy. In our dynamic model simulations, a decrease in the density of red blocks is represented by an increased number of self-avoiding beads, leading to higher local bead concentrations. This intensifies local crowding, strengthens repulsive forces, and ultimately raises the system's energy. Conversely, when the density of red blocks increases, they are represented by fewer beads, leading to lower local bead concentration and more available space. Thus, perturbations in bead number induce spatial heterogeneity in bead concentration among red blocks, which is accompanied by a rise in system energy.

Next, we analyzed how the system alleviates local crowding and dissipates excess energy following perturbations during random folding. In a crowded environment, regions of reduced bead concentration—arising from increased red block density—can serve as accessible sites for beads diffusing from more crowded regions, where red block density has decreased. The spatial clustering of red blocks into segregated conformations places regions of high and low bead concentration in close proximity, generating steeper concentration gradients that facilitate rapid bead diffusion, relieve local crowding, and accelerate the system's relaxation toward a lower-energy state. Indeed, we observed that the transition from partial to full red-blue segregation enables the system to alleviate local crowding more effectively. As shown by the averaged bead-pair distance distributions (Fig. 3E), in the late stage—when red and blue blocks were fully separated—relaxed conformations following perturbation exhibited a lower frequency of closely spaced bead pairs, indicative of reduced local crowding, compared to the partially segregated conformations observed in the early stage. These results confirm that, within a confined space, segregated conformations generate steeper bead concentration gradients, which improve energy dissipation efficiency following DNA-packing fluctuations, thereby energetically favoring their spontaneous formation during random folding.

We further investigated whether spatial confinement and sustained energy input from periodic perturbations are essential for initiating and maintaining segregation during random folding. Our folding simulations, conducted under periodic perturbations in spaces of different sizes, showed that as the available space increased, red-blue segregation gradually weakened and eventually became negligible (Fig. 4A-D). The absence of segregation can be explained as follows: in larger confining spaces, locally crowded regions—particularly those with decreased red block density—are surrounded by accessible spaces, allowing beads to diffuse away from these crowded regions. Consequently, the formation of a crowded region is coupled with the establishment of a steep bead concentration gradient relative to the surrounding accessible space, promoting rapid bead diffusion and facilitating energy dissipation. Under these conditions, the advantage of enhanced energy dissipation efficiency due to the spatial aggregation of red blocks is diminished, leading to weaker segregation between red and blue blocks. These results highlight the crucial role of spatial confinement in driving spontaneous segregation. Next, we investigated whether sustained energy input—delivered through periodic perturbations that keep the system out of thermodynamic equilibrium—plays a role in maintaining segregation, by conducting random folding simulations within a confined space. Starting from a fully segregated conformation, we removed all perturbations and allowed the system to relax. In the absence of periodic perturbations, segregation between red and blue blocks gradually diminished over time (Fig. 4E and Movie 6). These results suggest that sustained energy input is essential for preserving segregation. Together, the emergence and maintenance of the segregation depend on both spatial confinement and sustained energy input. Under these conditions, the spatially segregated organization arise through a non-equilibrium self-organizing process.

Finally, our results suggest that, within the constrained nuclear environment, fluctuations in DNA-packing density within euchromatin—associated with ATP-dependent chromatin remodelers—drive its spontaneous segregation from heterochromatin via a non-equilibrium self-organizing process. These findings offer a novel perspective for future models of genome organization.

**Acknowledgement**

We thank Prof. Wangjun Zhang (South China Normal University) for discussions. This work is supported by National Natural Science Foundation of China Youth Science Foundation Project (22403033), China Postdoctoral Science Foundation (2022M721203), and the Construction Plan of

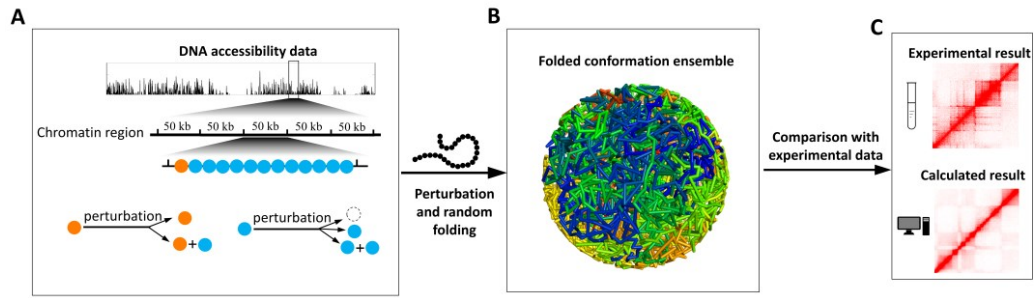

**Fig. 1. Polymer model to explore the effect of DNA-packing density fluctuations on 3D genome organization.**

(**A**) Schematic representation of chromatin polymer model. The chromatin fiber is partitioned into consecutive 50-kb segments, each modeled as a series of self-avoiding beads with equal physical volume. During random folding simulations, the number of beads per segment is periodically perturbed. In each perturbation, the first bead (orange) is constrained to either remain unchanged or split evenly into one orange and one cyan bead. The remaining beads (cyan) may disappear, remain unchanged, or divide evenly into two cyan beads. These perturbations dynamically alter the number of beads per segment, leading to fluctuations in the DNA-packing density of chromatin regions throughout the simulation. The model uses chromatin-accessibility data as the sole input.

(**B**) Ensemble of conformations sampled from folding simulation.

(**C**) Ensemble-averaged contact matrix generated for comparison with Hi-C maps. Details on periodical perturbation, folding simulations, conformation ensemble generation, and contact matrix construction are provided in the Methods section.

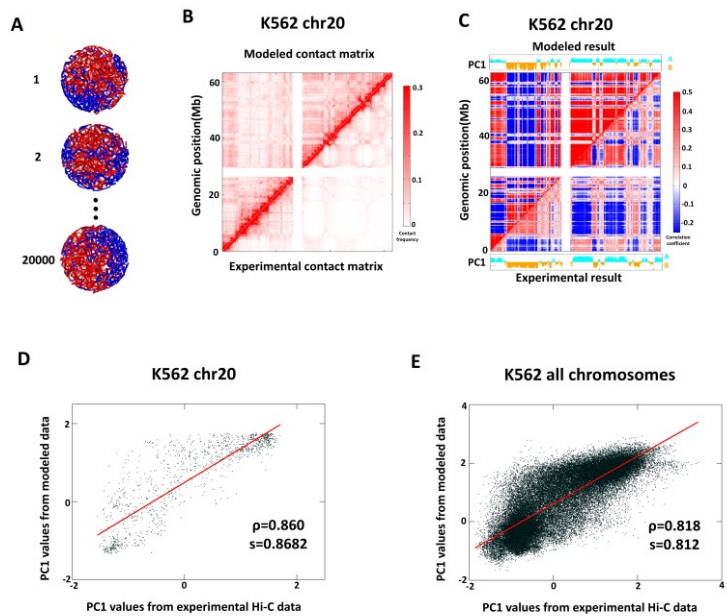

**Fig. 2. Comparison between simulated chromatin contacts and experimental Hi-C data.** For each autosome and the X chromosome of the K562 cell line, 20,000 conformations were sampled from folding simulations to generate conformation ensembles and their ensemble-averaged contact matrices at 50-kb resolution.

(**A**) Representative conformations sampled from the ensemble of 20,000 structures for chromosome 20, showing preferential spatial clustering of genomic regions assigned to the same compartment type (red for compartment A and blue for compartment B).

(**B**) Comparison of simulated contact matrix (upper triangle) and experimental Hi-C contact matrix (lower triangle) for chromosome 20.

(**C**) Comparison of correlation matrices derived from the contact matrices of simulated data (upper triangle) and experimental data (lower triangle) in **B**. Principal component 1 (PC1) values from two data sets are also shown. (**D** and **E**) Pearson correlations between simulated and experimental PC1 values for chromosome 20

(**D**) and for the entire genome, including all autosomes and the X chromosome

(**E**). The corresponding PC1 values used for these correlations are shown in Figure S3.

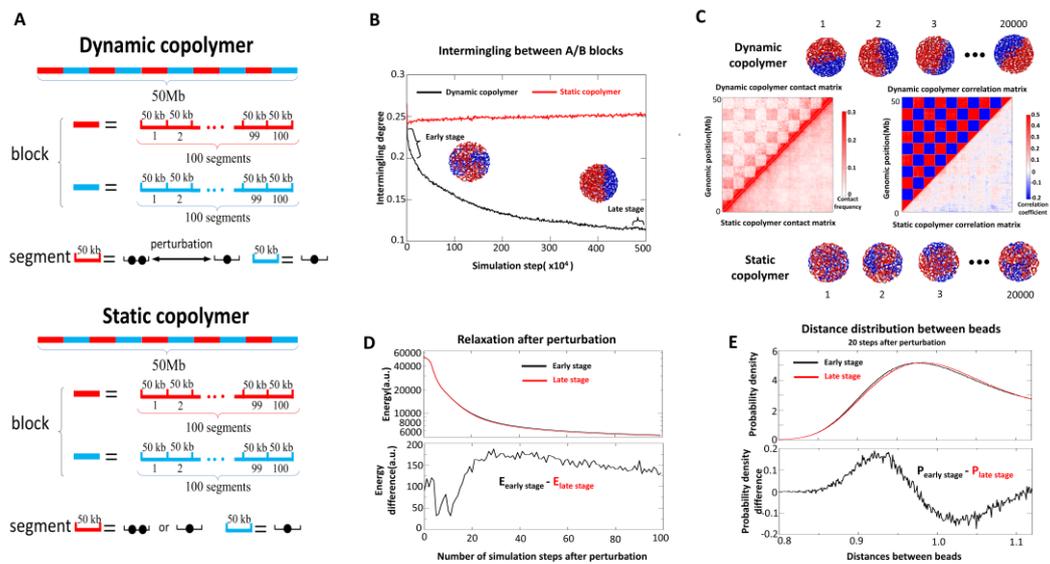

**Fig. 3. Fluctuations in DNA-packing density drive spatial segregation of euchromatin- and heterochromatin-like blocks in artificial chromatin fibers**

**(A)** Schematic of a 50-Mb artificial chromatin fiber composed of ten 5-Mb blocks, each containing 100 consecutive 50-kb segments. The blocks are categorized into two types: red euchromatin-like blocks, where each segment is represented by either one or two beads, and blue heterochromatin-like blocks, where each segment is consistently represented by a single bead. Chromatin conformations are generated using two simulation models. *Top:* In the **dynamic model**, red block segments undergo periodic perturbations during the folding simulation, allowing transitions between one-bead and two-bead representations. This perturbation simulates fluctuations in DNA-packing density. *Bottom:* In the **static model**, red block segment representations remain fixed—either one bead or two beads—with no perturbation. Simulated data from both models are used for subsequent analysis and comparison.

**(B)** Average intermingling degree between red and blue blocks over simulation steps for both models, computed from 100 independent trajectories. In the dynamic model, DNA-packing density fluctuations lead to pronounced segregation between block types. For downstream analyses, the first 100,000 steps are defined as the early stage, and the final 100,000 steps as the late stage (representative conformations shown).

**(C)** Comparison of sampled conformations, contact matrices, and correlation matrices from dynamic and static simulations.

**(D)** Average energy dissipation curves for the early and late stages defined in (B) (*top*) and their difference (*bottom*).

**(E)** Average bead-pair distance distributions for early- and late-stage conformations, sampled at the 20th step after perturbation (*top*), and the difference between them (*bottom*).

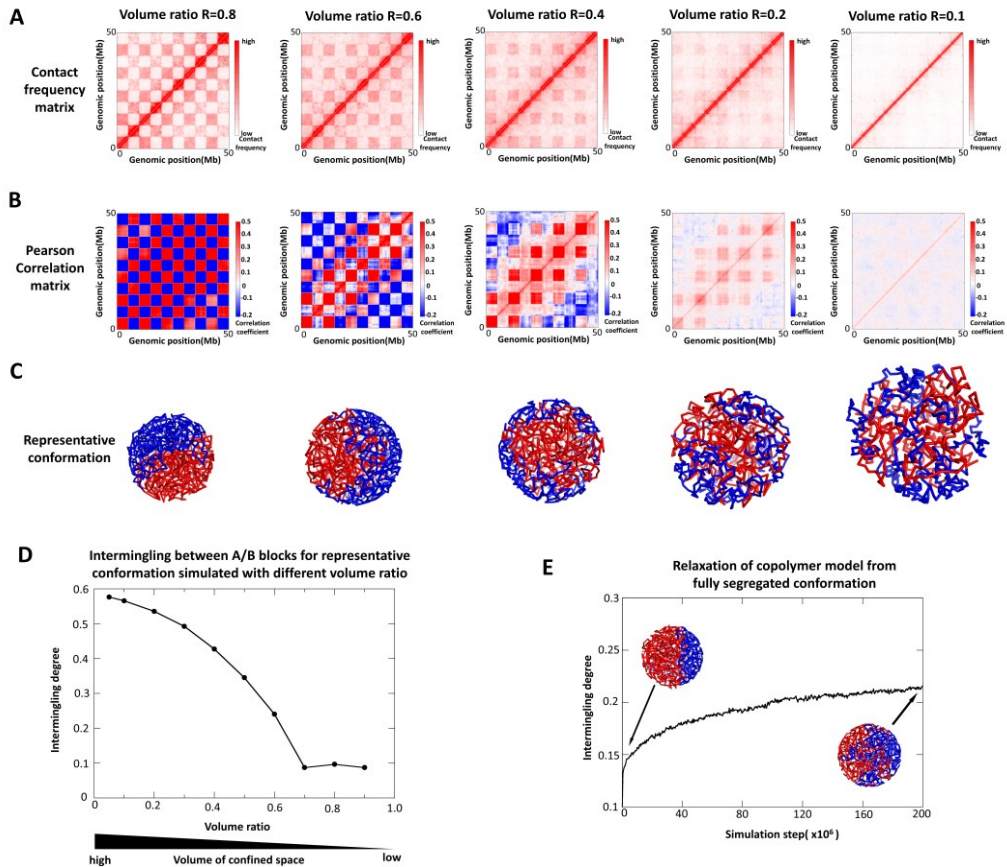

**Fig. 4. Emergence and maintenance of euchromatin–heterochromatin segregation require spatial confinement and sustained fluctuations in DNA-packing density.**

**(A–C)** Ensemble-averaged contact matrices (A), correlation matrices (B), and representative conformations (C) from random folding simulations of the 50-Mb artificial chromatin fiber using the dynamic model (as described in Fig. 3), under varying degrees of spatial confinement. The degree of confinement is quantified by the volume occupancy ratio $R$, defined as the ratio of the total volume of all polymer beads to the volume of the confining sphere.

**(D)** Average intermingling degree between red and blue blocks in final conformations, plotted as a function of volume occupancy ratio $R$. For each $R$ value, the intermingling degree is computed from the final conformations of 100 independent folding trajectories.

**(E)** Evolution of the intermingling degree between red and blue blocks over simulation steps in the static model, starting from segregated conformations. The analysis is based on 100 independent trajectories, each initialized from the final conformations obtained in the dynamic model (Fig. 3B). Representative initial and final conformations are shown.

# Supplementary Materials for

**Fluctuations in DNA Packing Density Drive the Spatial Segregation between Euchromatin and Heterochromatin**

Luming Meng, Boping Liu and Qiong Luo

Corresponding authors: Luming Meng, menglum@scau.edu.cn; Qiong Luo, luoqiong@scnu.edu.cn

**The file includes:**

   Methods
   Figs. S1 to S6
   References

**Other Supplementary Material for this manuscript includes the following:**

   Captions of Movies 1-6

# Context
**Methods**

    I. The Polymer Model for Investigating the Effect of DNA-Packing Density Fluctuations on 3D Genome Organization

    II. Random Folding Simulation

    III. Conformation Ensembles Generated in This Paper

    IV. Artificial Chromatin Fiber

    V. Constructing contact matrices from conformation ensembles

    VI. Identification of A/B compartments

    VII. The method for the identification of A/B compartments from Hi-C contact matrix with very sparse data

    VIII. Filtering ATAC-seq signals of the mouse 8-cell and ICM stages

    IX. Source list of experimental data

**Supplementary Figures**

**Captions of Movies**

**References**

**Methods**

**I. The Polymer Model for Investigating the Effect of DNA-Packing Density Fluctuations on 3D Genome Organization**

We first partition the chromatin region of interest into $N$ consecutive 50-kb segments. Each segment is modeled as $k$ contiguous, self-avoiding coarse-grained beads, thereby generating a polymer chain representing the chromatin region. The DNA-packing density of each 50-kb segment is defined as inversely proportional to the number of beads it contains, *i.e.*, proportional to $50/k$. For clarity in the following description, beads within each segment are distinguished by color: the first bead is orange, while the others are cyan (Fig. 1A). We simulate the random folding of this polymer chain in a confined space over 2,000,000 steps of numerical integration (see Section II for simulation details). To incorporate fluctuations in DNA-packing density during the folding simulation, we update the bead number ($k$) for each 50-kb segment every 100 simulation steps, using population-averaged chromatin accessibility data derived from DNase-seq or ATAC-seq experiments. The procedure for updating the bead number is detailed below:

(1) *Initial Representation of 50-kb Segments*:

For simplicity, our model initially represents each 50-kb segment by a single bead, which corresponds to the first bead of the segment and is colored orange. Thus, the chromatin region of interest is initially modeled as a self-avoiding polymer chain composed of $N$ beads, all marked in orange.

(2) *Assignment of DNA Accessibility Signals for Each Bead:*

For each bead, the DNA accessibility signal $\zeta$ is determined by mapping the experimental population-averaged chromatin accessibility data (at 1-kb resolution) onto the genomic sequence corresponding to that bead. The accessibility signal $\zeta_i$ for the $i^{th}$ bead is defined as the total number of reads mapped to its corresponding sequence. The normalized DNA accessibility signal $\lambda_i$ for the $i^{th}$ bead is calculated as follows:

$$\lambda_i = \frac{\zeta_i}{2\bar{\zeta}} \qquad (1)$$

where $\bar{\zeta}$ is the average DNA accessibility signal across all beads, defined as

$$\bar{\zeta} = \frac{1}{N}\sum_{i=1}^{N}\zeta_i \qquad (2)$$

where *N* is the total number of beads in the polymer chain of the chromatin region of interest.

(3) *Periodic Perturbations in the Number of Beads per Segment during Folding Simulation*

As described above, the chromatin region of interest is initially represented as a self-avoiding polymer chain consisting of *N* beads, with each 50-kb segment represented by a single orange bead (*i.e.*, one bead per segment). A random initial conformation of this chain is generated using a self-avoiding random walk algorithm. Starting from this conformation, the random folding of the chromatin region is simulated over 2,000,000 numerical integration steps. Throughout the simulation, perturbations to the bead number (*k*) in each 50-kb segment are applied every 100 steps. The details of the periodic perturbations are outlined below:

(i) <u>The First Perturbation:</u>

After the first 100 steps of numerical integration, we perturb the number of beads for all 50-kb segments, based on the normalized DNA accessibility signals of beads. Specifically, we calculate the normalized DNA accessibility signal $\lambda$ for each orange bead in the conformation obtained after the first 100 steps of numerical integration. Each orange bead has two possible fates:

*No Change*: remain unchanged with a probability of $\frac{1}{2} + \frac{1}{2} max\{1 - \lambda, 0\}$.

*Splitting*: evenly divide into one orange and one cyan bead with a probability of $\frac{1}{2} min\{\lambda, 1\}$.

Notably, "evenly divide" means that the original orange bead splits into two new beads (one orange bead and one cyan bead), each retaining the same physical volume as the original but representing half of its genomic sequence. Throughout this study, all instances of 'evenly divide' refer to this definition. The newly generated orange bead always precedes the cyan bead to ensure that the first bead in each segment remains orange. The coordinates of the two newly generated beads are determined using an interpolation algorithm, as detailed in Section II. This perturbation produces a new conformation with a revised DNA-packing density distribution along the chromatin region of interest.

(ii) <u>Perturbations After the First Perturbation:</u>

After the first perturbation, each 50-kb segment may comprise multiple beads, with the first being an orange bead followed by cyan beads (Fig. 1A). At each perturbation step, we always first calculate the normalized accessibility signal ($\lambda$) for each bead and then determine its fate based on its $\lambda$ value.

The orange bead has two possible fates as before:

*No Change*: remain unchanged with a probability of $\frac{1}{2} + \frac{1}{2}max\{1 - \lambda, 0\}$.

*Splitting*: evenly divide into one orange and one cyan bead with a probability of $\frac{1}{2}min\{\lambda, 1\}$.

As before, the newly generated orange bead always precedes the cyan bead to ensure that the first bead in each segment remains orange. Coordinates of all newly generated beads are determined using the interpolation algorithm (see Section II below).

Each pre-existing cyan bead within a segment—excluding the newly generated cyan bead resulting from the splitting of an orange bead—can follow one of three possible fates:

*No Change*: remain unchanged with a probability of $\frac{1}{2}$.

*Splitting*: evenly divide into two cyan beads with a probability of $\frac{1}{2}min\{\lambda, 1\}$.

*Disappearance*: disappear with a probability of $\frac{1}{2}max\{1 - \lambda, 0\}$.

If a bead disappears, its genomic sequence is assigned to the preceding bead to maintain continuity. Coordinates of all newly generated beads are determined using the interpolation algorithm (see Section II below).

Overall, the cycle of 100-step simulations followed by bead number perturbations is repeated 20,000 times, resulting in 2,000,000 numerical integration steps and 20,000 perturbation events. This procedure dynamically adjusts the bead number ($k$) for each 50-kb segment, thereby introducing fluctuations in local DNA-packing density (i.e., 50/$k$) during folding and generating a random folding trajectory for the chromatin region of interest. To account for variability due to initial conditions, the entire simulation process is performed 100 times, each starting from a distinct initial conformation of the polymer composed of $N$ orange beads. This yields 100 independent folding trajectories, representing possible chromatin folding pathways. Conformations are saved every 10,000 steps from each trajectory, resulting in an ensemble of 20,000 conformations for analyzing how local DNA-packing fluctuations influence chromatin folding.

**II. Random Folding Simulation**

This section outlines the numerical integration procedure and the algorithm used to assign coordinates to newly generated beads following bead number perturbations.

(1) *Numerical Integration*

Throughout each 100-step numerical integration, the number of beads used to represent the chromatin region remains unchanged. Each bead is characterized by a diameter $b$ and mass $m$. The step size for the integration, $\Delta t$, is defined as:

$$\Delta t = \sqrt{0.2b * \frac{m}{F_{max}}} \qquad (3)$$

where $F_{max}$ is the maximum force among the forces applied on beads in the polymer. The diameter $b$ and the mass $m$ of each bead are set to 1 arbitrary unit. The conformation of the polymer is updated with the formula:

$$\vec{r}_{i,t+\Delta t} = \vec{r}_{i,t} + \vec{v}_{i,t}\Delta t + \frac{\vec{f}_{i,t}}{2m}\Delta t^2 \qquad (4)$$

where $\vec{r}_{i,t}$ and $\vec{r}_{i,t+\Delta t}$ separately represent the coordinates of the $i^{th}$ bead before and after the integration step. The velocity vectors $\vec{v}_{i,t}$ are drawn from the Gaussian distribution at every integration step. $\vec{f}_{i,t}$ represents the force applied on the $i^{th}$ bead based on the potential energy described below. $m$ is the mass of each bead with the value of 1 and $\Delta t$ is the step size defined in Eq. 3.

Potential energy used in numerical integration. At each step of the numerical integration, the total potential energy of a chromatin conformation is defined as the sum of three components, detailed below:

(i) Harmonic oscillator potential energy, $E_{har}(r_{n,n+1})$: This term maintains the connectivity of the polymer by modeling the interaction between sequentially adjacent beads as harmonic springs. It is computed as:

$$E_{har}(r_{n,n+1}) = \frac{1}{2}k_b(r_{n,n+1} - b)^2 \qquad (5)$$

where $r_{n,n+1}$ is the Euclidean distance between the sequentially adjacent $n^{th}$ and $(n+1)^{th}$ beads. The equilibrium distance between the $n^{th}$ and $(n+1)^{th}$ beads is equal to the bead diameter $b$ with the value of 1 arbitrary unit. The force constant $k_b$ is defined as:

$$k_b = \frac{200k_BT}{b^2} \qquad (6)$$

where $k_B$ is the Boltzmann constant, $T$ the temperature, and $b$ the diameter of bead with the value of 1 arbitrary unit. Here, $k_BT$ is considered as a unit amount of energy with value of 1 arbitrary unit.

(ii) Repulsive potential energy, $E_{WCA}(r_{n,m})$: This term prevents spatial overlap between non-sequentially adjacent beads and is computed as:

$$E_{WCA}(r_{n,m}) = \begin{cases} 4\varepsilon\left(\left(\frac{\sigma}{r_{n,m}}\right)^{12} - \left(\frac{\sigma}{r_{n,m}}\right)^{6}\right), & if\ r_{n,m} \leq 2^{1/6}\sigma \\ -\varepsilon, & else \end{cases} \qquad (7)$$

where $r_{n,m}$ represents the Euclidean distance between the $n^{th}$ and the $m^{th}$ beads and $\sigma$ and $\varepsilon$ are the two constants with $\sigma = b$ and $\varepsilon = \frac{1}{2}k_BT$.

(iii) Spherical restraint potential energy, $E_{sphere}(r_n)$: This term confines the polymer within a spherical space, mimicking the spatial constraints of a cell nucleus. The center of the sphere is fixed at the origin (0, 0, 0) throughout the simulation. The radius of the sphere $r_s$ is defined as:

$$r_s = \frac{1}{2}b((N + \Lambda)/0.74)^{1/3} \qquad (8)$$

where $b$ is the bead diameter, set to 1 arbitrary unit; $N$ is the total number of 50-kb segments; and $\Lambda$ denotes the sum of the normalized DNA accessibility signals of $N$ segments. Given this definition of the confinement space, the spherical restraint potential energy, $E_{sphere}(r_n)$, is calculated as:

$$E_{sphere}(r_n) = \begin{cases} \frac{r_n - r_s}{b}kT, & if\ r_n \geq r_s \\ 0, & else \end{cases} \qquad (9)$$

where $r_n$ is the Euclidean distance between the $n^{th}$ bead and the center of the sphere space (i.e. the origin of the coordinate system).

(2) *Determination of the coordinates of newly generated beads following bead number perturbations*

At each perturbation step, the bead number within each 50-kb segment is updated. Some beads in the relaxed conformation from the preceding 100-step integration are evenly divided into two new beads. To initiate the next 100-step integration, spatial coordinates must be assigned to these newly generated beads. Specifically, if the $i^{th}$ bead is divided into two new beads, the coordinates of the $i^{th}$ and $i+1^{th}$ beads from the previous relaxed conformation, denoted as $\vec{r}_{i,pre}$ and $\vec{r}_{i+1,pre}$, are used to calculate the coordinates of the newly generated beads, $\vec{r}'_{i,new}$ and $\vec{r}''_{i,new}$, according to

the following:

$$\vec{r}'_{i,new} = \vec{r}_{i,pre} \tag{10}$$

$$\vec{r}''_{i,new} = \frac{(\vec{r}_{i,pre} + \vec{r}_{i+1,pre})}{2} \tag{11}$$

**III. Conformation Ensembles Generated in This Paper**

As described in Sections I and II, the chromatin region of interest is partitioned into $N$ consecutive 50-kb segments, each initially represented by a single bead, forming a polymer chain of $N$ beads. Starting from a randomly generated initial conformation of this polymer chain, we perform a random folding simulation over 2,000,000 numerical integration steps. During the simulation, the bead number of each 50-kb segment is perturbed every 100 steps, and conformations are saved every 10,000 steps. This simulation is repeated 100 times from distinct initial conformations, producing 100 independent folding trajectories. In total, 20,000 conformations are collected to form an ensemble for investigating the effect of DNA-packing density fluctuations on chromatin folding behavior.

(1) *Conformation ensembles for K562 human erythroleukemia cell line*

For the chromatin region of interest spanning the 22 autosomes and the X chromosome in the K562 cell line, we generated conformation ensembles for each chromosome using population-averaged ATAC-seq data (*32*). The results of the analysis on these ensembles are presented in Fig. 2 and Figure S3. The source code used to generate these conformation ensembles—using the K562 population-averaged ATAC-seq dataset as an example—is available at https://github.com/TheMengLab/chromosome_3D_phase_separation_structures_from_DNA_accessibility .

(2) *Conformation ensembles of the cell types involved in early mammalian development*

Using chromatin accessibility data from Du et al.'s study on preimplantation mouse development (*34*), we generated chromatin conformation ensembles for all 19 autosomes at both the 8-cell and inner cell mass (ICM) stages. The results on these ensembles are presented in Figure S1, S2, S4, and S5.

**IV. Artificial Chromatin Fiber**

The 50-Mb artificial chromatin fiber comprises 1,000 consecutive 50-kb segments, organized into

10 alternating euchromatin-like (red) and heterochromatin-like (blue) blocks, each containing 100 segments (Fig. 3a). We constructed two models: a *static copolymer*, in which the DNA-packing density of all blocks remained fixed throughout the folding simulation, and *a dynamic copolymer*, in which the density of red euchromatin-like blocks fluctuated over time.

(1) *The Static Model*

In the static model, each segment within both the euchromatin-like (red) and heterochromatin-like (blue) blocks was assigned a fixed number of beads. Specifically, of the 500 red segments, 250 were randomly selected to contain two beads each, while the remaining 250 red segments and all 500 blue segments each contained one bead, resulting in a total of 1,250 beads for the 50-Mb artificial chromatin fiber. In this setup, blue heterochromatin-like blocks had a density of 50 kb per bead (i.e., 5 Mb per 100 beads), while red euchromatin-like blocks had lower densities ranging from 25 kb or 50 kb per bead. The fiber was then randomly folded within a confined space for 5,000,000 integration steps, with the bead number in each segment held constant to maintain a fixed DNA-packing density throughout the folding process and prevent density fluctuations. This simulation was repeated 100 times with distinct initial conformations, generating 100 independent folding trajectories. Conformations were saved every 10,000 steps, resulting in 500 structures per trajectory. Among these, the last 200 conformations from each trajectory were retained for constructing contact matrices, yielding an ensemble of 20,000 conformations. For each trajectory, the degree of separation between the red and blue blocks was quantified across 500 sampled conformations and averaged across all trajectories to obtain the average separation degree (Fig. 3).

(2) *The dynamic Model*

The dynamic model used the same 50-Mb, 1,250-bead chromatin fiber as the static model. However, in this model, the number of beads per segment in the red euchromatin-like blocks fluctuated during the folding simulation, while each segment in the blue heterochromatin-like blocks consistently contained a single bead. Initially, 250 of the 500 red segments were randomly assigned two beads each, while the remaining 250 red segments contained one bead each. Every 100 steps during the 5,000,000-step folding simulation, 250 red segments were randomly selected to contain two beads, while the remaining 250 segments contained one bead. This fluctuation allowed each of the 500 red segments to alternate between one and two beads multiple times during the simulation, while maintaining a constant total bead count for the 50-Mb fiber. As in the static model, the simulation was repeated 100 times with distinct initial conformations, generating 100 independent folding trajectories. Conformations were sampled every 10,000 steps, yielding 500 structures per trajectory. Among these simulated conformations, the last 200 from each trajectory were retained for constructing contact matrices, resulting in an ensemble of 20,000 conformations. For each trajectory, the degree of separation between the red and blue blocks was quantified across the 500 sampled conformations and averaged across all trajectories to obtain the average separation degree between the two blocks (Fig. 3).

(3) *The Degree of Intermingling between the Red and Blue Blocks*

To quantify the degree of separation between euchromatin-like (red) blocks and heterochromatin-like (blue) blocks in a given conformation, we defined *I* as a measure of the intermingling degree between beads within both types of blocks:

$$I = \frac{1}{N_{bead}} \sum_{i=1}^{N_{bead}} I_i \quad (12)$$

where $N_{bead}$ is the total number of beads in the chromatin fiber, and $I_i$ represents the intermingling degree for the $i^{th}$ bead, defined as

$$I_i = \frac{D_i}{S_i + D_i} \quad (13)$$

where $S_i$ is the number of beads that share the same color as the $i^{th}$ bead and are located within the spatial distance of less than 1 bead diameter from the $i^{th}$ bead, while $D_i$ is the number of beads with a different color located within 1-bead-diameter space of the $i^{th}$ bead.

(4) *Folding Simulations with Periodic Perturbations in Spaces of Different Sizes*

We used the dynamic model to perform folding simulations in spheres of different sizes to investigate the role of spatial confinement in the folding behavior of the artificial chromatin fiber. The radius of the sphere confining the artificial chromatin fiber is defined as:

$$r_s = \frac{1}{2} b (N_{bead}/R)^{1/3} \quad (14)$$

where *b* is the bead diameter with the value of 1 arbitrary unit, and $N_{bead}$ is the total number of beads used to represent the chromatin fiber. *R* is the volume occupancy rate, defined as the ratio between the total volume of all polymer beads and the spatial volume of the sphere, with a default value of 0.74. For simulations investigating the role of spatial confinement, the value of *R* is varied from 0.1 to 0.9 with increments of 0.1.

For each value of *R*, we performed 100 independent folding simulations using the dynamic model, each initiated from a distinct random conformation, yielding 100 folding trajectories. From each trajectory, conformations were sampled every 10,000 steps, resulting in 500 structures per trajectory. The last 200 conformations from each trajectory were selected to construct contact matrices, generating an ensemble of 20,000 conformations. Representative contact maps are shown in Fig. 4. For each trajectory, the intermingling degree between the red and blue blocks in the final conformation was quantified, and the values were averaged across all trajectories to obtain the mean separation degree as a function of *R*, as shown in Fig. 4D.

(5) *Folding Simulation from a Fully Segregated Conformation in a Confined Space without Periodic Perturbations*

To evaluate whether sustained energy input from periodic perturbations is required to initiate and maintain segregation between red and blue blocks during random folding, we applied the static model to simulate the folding of the artificial chromatin fiber within a confined sphere, with the volume occupancy rate *R* in Eq. (14) set to 0.74. A total of 100 trajectories were generated, each starting from a fully segregated conformation obtained using the dynamic model. Simulations were run for 200 million integration steps per trajectory, following the procedures described above for the static model. Conformations were saved every 400,000 steps, yielding 500 structures per trajectory. From these, the final 100 conformations of each trajectory were retained, resulting in an ensemble of 10,000 conformations for downstream analysis. The degree of separation between red and blue blocks was quantified across all 500 sampled conformations per trajectory and then averaged across the 100 trajectories to obtain the mean intermingling degree (Fig. 4E).

The code for simulating artificial 50-Mb chromatin fiber is deposited at https://github.com/TheMengLab/artificial_chromosome_phase_separation_mechanism .

**V Constructing contact matrices from conformation ensembles**

For all the conformation ensembles mentioned above, we construct the following:

(1) *Individual contact matrices for each conformation*

<u>Individual bead-level contact matrices for each conformation ($C^{bead}$)</u>. For each conformation, the Euclidean distance between every pair of beads is measured. Based on the measurement, we build a bead-level contact matrix $C^{bead}$ for each conformation, where the matrix element $c_{ij}^{bead} = 1$ if the distance between the $i^{th}$ and $j^{th}$ beads is less than 1 bead diameters and $c_{ij}^{bead} = 0$ otherwise.

Individual segment-level contact matrices ($C^{seg}$). The individual segment-level contact matrix, $C^{seg}$, is derived from $C^{bead}$. Beads $i$ and $j$ in $C^{bead}$ are mapped to their corresponding 50-kb genomic segments. The matrix element $c_{mn}^{seg}$ in $C^{seg}$ is determined by the number of contacts between beads belonging to the $m^{th}$ and $n^{th}$ segments. This process enables the transition from individual bead-level interactions to biologically meaningful segment-level contact matrices, facilitating analysis at a 50-kb resolution.

(2) *Normalized ensemble-average contact matrix* $\tilde{A}$

For each ensemble, we compute the ensemble-averaged contact matrix $A$ by averaging the individual segment-level contact matrices $C^{seg}$ across all conformations. Specifically, $A$ is defined as:

$$A_{ij} = \frac{1}{N}\sum_{n=1}^{N} c_{ij,n} \qquad (12)$$

where $N$ is the total number of conformations in the ensemble, and $c_{ij,n}$ denotes the contact count between the $i^{th}$ and $j^{th}$ segments in the $n^{th}$ conformation.

The normalized ensemble-average contact matrix $\tilde{A}$ is derived from $A$ using the K-R algorithm(*36*). This normalization adjusts the matrix so that the row and column sums of $\tilde{A}$ are equal, ensuring that each bin has an equal total number of interactions with other bins. Normalization serves two key purposes. First, it removes biases introduced by differences in raw contact counts, enabling a more accurate comparison of interaction patterns. Second, it preserves the relative distribution of contacts within rows and columns, facilitating the identification of structural features such as regions of high interaction density. By rescaling the matrix in this way, $\tilde{A}$ emphasizes relative interaction patterns rather than absolute frequencies, allowing for more meaningful analyses of chromatin organization.

The code for generating contact matrix from a conformation ensemble is deposited at https://github.com/TheMengLab/chromosome_3D_phase_separation_structures_from_DNA_accessibility .

**VI. Identification of A/B compartments**

The identification of A/B compartment structures from both the population-averaged Hi-C contact matrix and the normalized ensemble-average contact matrix $\tilde{A}$ follows the algorithm described in previous work(*31*). Below, we present the steps for identifying compartments using the Hi-C contact-frequency matrix as an example.

(1) *Normalization of the Hi-C matrix*

Each matrix element is divided by the genome-wide average contact frequency for bin pairs at the same genomic distance, producing the normalized Hi-C matrix.

(2) *Calculation of the correlation matrix $P$*

A correlation matrix $P$ is constructed, where each element $P_{ij}$ represents the Pearson correlation between the $i^{th}$ and $j^{th}$ rows of the normalized Hi-C matrix. This correlation matrix reveals large blocks of enriched and depleted interactions, forming plaid-like patterns.

(3) *Principal component analysis (PCA)*

To assign bins to compartments, PCA is applied to the correlation matrix $P$, as described in previous work (*10*). Specifically: The matrix $P$ is decomposed to obtain eigenvalues and eigenvectors. The eigenvector corresponding to the largest eigenvalue is designated as the first principal component (PC1). Genomic bins can be classified into two compartments based on the sign of their PC1 values. Bins in the A compartment are characterized by higher DNA accessibility signals. If the A compartment corresponds to a negative PC1 value, the entire PC1 vector should be multiplied by -1 to ensure consistent interpretation. Genomic bins are categorized based on the sign of their PC1 values :

  <u>Positive PC1 values</u>: Indicate A compartments, which are euchromatic, highly accessible, and associated with active transcription.

  <u>Negative PC1 values</u>: Indicate B compartments, characterized as heterochromatic, less accessible, and linked to transcriptional inactivity.

## VII. The method for identifying A/B compartment from Hi-C contact matrix with very sparse data

In early mammalian development, the number of available cells is highly limited, as shown in studies of preimplantation mouse embryos (*34*). In these developmental stages, Hi-C experiments often yield smaller datasets, with contacts primarily occurring between bins that are in close genomic proximity (Figure S1B and S2B). As a result, directly applying the method described in Section V to assign bins to A/B compartments using sparse Hi-C maps becomes challenging.

  To address this issue, we developed a method specifically designed to identify A/B compartments from Hi-C contact matrices with very sparse data. As an example, we outline the steps of this method using the Hi-C matrix of chromosome 19 from the 8-cell stage of mammalian development (Figure S1). The method includes the following steps:

(1) *Normalization of the Hi-C Matrix*

We first normalize the Hi-C matrix according to the procedure described in Section V. This results in a normalized matrix where each contact frequency is adjusted based on the genome-wide average for bin pairs at the same genomic distance.

(2) *Cutting Overlapping Squares Along the Diagonal*

Next, we cut a series of overlapping squares along the diagonal of the normalized Hi-C map. Each square represents a sub-set of the chromatin interaction frequency map, with the size of the square corresponding to a predefined genomic region length, such as 10 Mb.

(3) *General Case for Overlapping squares*

Consider a genomic region of length $L$, covered by $N$ overlapping squares, each representing a segment consisting of $M$ bins. The segments are defined as:

- Square #1: Bins 1 to $M$.
- Square #2: Bins $k+1$ to $M+k$.
- Square #3: Bins $2k+1$ to $M+2k$.
- ...
- Square #$N$: Bins $(N-1)k+1$ to $M+(N-1)k$.

Here, $k$ is the step size ($k<M$). This setup ensures that the genomic region is divided into overlapping segments, each of which spans $M$ bins and overlaps with its neighbors by $M-k$ bins. The total region length $L$, number of segments $N$, and overlap $M-k$ can be adjusted based on the data or analysis needs.

(4) *Assigning bins to A/B compartments (example : a 20-Mb Region)*

As an example, consider a 20-Mb ($L=20$ Mb) region covered by three overlapping squares of 10 Mb size with the step size of 5 Mb and bin size of 50 kb ($N=400$, $M=200$, $k=100$):

- Square **a**: Bins 1–200.
- Square **b**: Bins 100–300.
- Square **c**: Bins 200–400.

For each square, we calculate its correlation matrix following the approach described in Section V. The principal component analysis (PCA) is then applied to each correlation matrix to extract the eigenvector corresponding to the largest eigenvalue, denoted as the first principal component (PC1). For the 2-Mb example, the PC1 vectors for squares **a**, **b**, and **c** are:

$\vec{V_a} = (v_{a1}, v_{a2}, \cdots, v_{a200})$, where $v_{ai}$ $(i = 1 - 200)$ represents the value of PC1 of the $i^{th}$ bin.

$\vec{V_b} = (v_{b100}, v_{b101}, \cdots, v_{b300})$, where $v_{bi}$ $(i = 100 - 300)$ represents the value of PC1 of the $i^{th}$ bin.

$\vec{V_c} = (v_{c200}, v_{c201}, \cdots, v_{c400})$, where $v_{ci}$ $(i = 200 - 400)$ represents the value of PC1 of the $i^{th}$ bin.

To assign bins in the entire 20-Mb region to A/B compartments, we need determine the PC1 vector $\vec{X}(x_1, x_2, \cdots, x_{400})$ of the entire region. Here, $\vec{X}$ is determined by minimizing the differences between $\vec{X}$ and the scaled PC1 vectors of overlapping squares:

$$min\left\{d\{c_a\vec{V_a}, \vec{X}\} + d\{c_b\vec{V_b}, \vec{X}\} + d\{c_c\vec{V_c}, \vec{X}\}\right\}$$

$$d\{c_a\vec{V_a}, \vec{X}\} = \sum_{i=1}^{200}(c_a v_{ai} - x_i)^2,$$

$$d\{c_b\vec{V_b}, \vec{X}\} = \sum_{i=100}^{300}(c_b v_{bi} - x_i)^2,$$

$$d\{c_c\vec{V_c}, \vec{X}\} = \sum_{i=200}^{400}(c_c v_{ci} - x_i)^2$$

where $c_a$, $c_b$, and $c_c$ are scaling factors.

(5) *The validation of the method*

To validate the method, we applied this algorithm to calculate the compartment PC1 values for K562 chromosome 10 based on Hi-C contact reads with genomic distance less than 50 Mb. The parameters of PC1 calculation are binsize=50 kb, *N*=2712, *M*=1000 and *k*=167. For comparison, we also used the method described in Section VI to generate the compartment PC1 values using all Hi-C contact reads. Then we compare the profiles of PC1 values from two approaches to validate this algorithm. The result shows that the profiles of the two PC1 values are highly correlated, with a correlation coefficient of 0.956 (shown in Figure S6).

The source code for identifying A/B compartment from experimental Hi-C contact matrix of sparse read is deposited at https://github.com/TheMengLab/compartment_from_sparse_Hi-C .

**VIII. Filtering ATAC-seq signals at the mouse 8-cell and ICM stages**

To minimize the influence of noise signals on the construction of chromatin structures at the mouse 8-cell and ICM stages, ATAC-seq signals are filtered. First, the chromatin is divided into 1-kb bins, and the ATAC-seq signals for each bin are computed based on experimental data. The top 10% of bins with the highest signals are classified as target regions, while the remaining 90% are treated as noise. In other words, only the signals from the top 10% of genomic regions are retained, and the signals from the remaining 90% are discarded for constructing chromatin structures at these stages.

## IX Source list of experimental data

| REAGENT or RESOURCE | SOURCE | IDENTIFIER |
|---|---|---|
| Software and algorithms | | |
| Code for generating and analyzing conformation ensembles | This paper | https://github.com/TheMengLab/chromosome_3D_phase_separation_structures_from_DNA_accessibility<br>https://github.com/TheMengLab/artificial_chromosome_phase_separation_mechanism<br>https://github.com/TheMengLab/compartment_from_sparse_Hi-C |
| Data source | | |
| Population averaged Hi-C data for K562 | S. S. P. Rao *et al.*, A 3D Map of the Human Genome at Kilobase Resolution Reveals Principles of Chromatin Looping (vol 159, pg 1665, 2014). *Cell* **162**, 687-688 (2015). | GEO:GSE63525 |
| Population averaged ATAC-seq data for K562 | | ENCODE:ENCSR483RKN |
| Hi-C data for mouse 8-cell and ICM | Z. H. Du *et al.*, Allelic reprogramming of 3D chromatin architecture during early mammalian development. *Nature* **547**, 232-+ (2017). | GSE82185 |
| ATAC-seq for mouse 8-cell and ICM | J. Y. Wu *et al.*, The landscape of accessible chromatin in mammalian preimplantation embryos. *Nature* **534**, 652-+ (2016). | GSE66581 |

**Supplementary Figures**

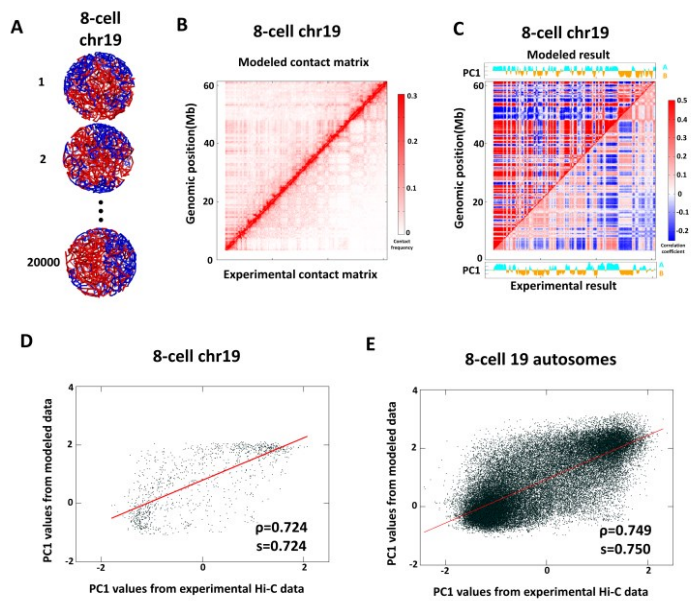

**Figure S1. Comparison between simulated chromatin contacts and experimental Hi-C data for mouse 8-cell.** For 19 autosomes at the 8-cell stage of mouse, 20,000 conformations were sampled from folding simulations to generate conformation ensembles and their ensemble-averaged contact matrices at 50-kb resolution. (**A**) Representative conformations sampled from the ensemble of 20,000 structures for chromosome 19, showing preferential spatial clustering of genomic regions assigned to the same compartment type (red for compartment A and blue for compartment B). (**B**) Comparison of simulated contact matrix(upper triangle) and experimental Hi-C contact matrix(lower triangle) for chromosome 19. (**C**) Comparison of correlation matrices derived from the contact matrices of simulated data(upper triangle) and experimental data(lower triangle) in **B**. Principal component 1 (PC1) values from two data sets are also shown. (**D** and **E**) Pearson correlations between simulated and experimental PC1 values for chromosome 19 (**D**) and for the entire genome, including 19 autosomes (**E**). The corresponding PC1 values used for these correlations are shown in Figure S4.

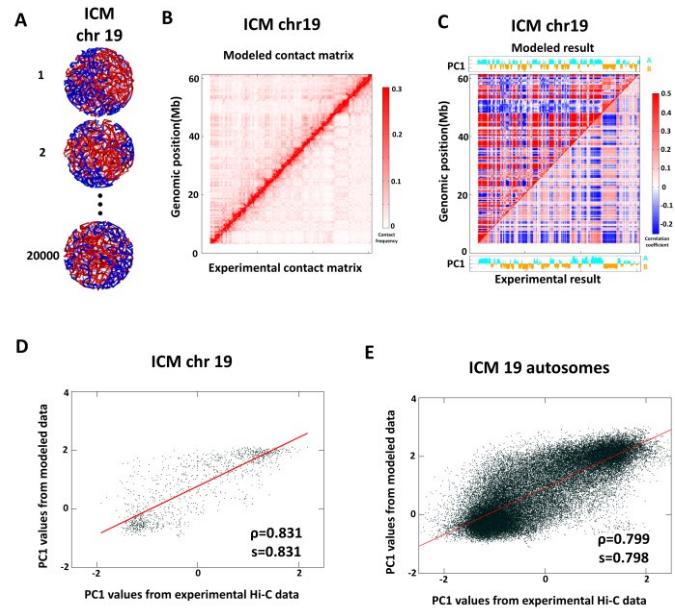

**Figure S2. Comparison between simulated chromatin contacts and experimental Hi-C data for mouse ICM cells.** For each autosome at the ICM stage of mouse, 20,000 conformations were sampled from folding simulations to generate conformation ensembles and their ensemble-averaged contact matrices at 50-kb resolution. **(A)** Representative conformations sampled from the ensemble of 20,000 structures for chromosome 19, showing preferential spatial clustering of genomic regions assigned to the same compartment type (red for compartment A and blue for compartment B). **(B)** Comparison of simulated contact matrix(upper triangle) and experimental Hi-C contact matrix(lower triangle) for chromosome 19. **(C)** Comparison of correlation matrices derived from the contact matrices of simulated data(upper triangle) and experimental data(lower triangle) in **B**. Principal component 1 (PC1) values from two data sets are also shown. (**D** and **E**) Pearson correlations between simulated and experimental PC1 values for chromosome 19 (**D**) and for the entire genome, including 19 autosomes (**E**). The corresponding PC1 values used for these correlations are shown in Figure S5.

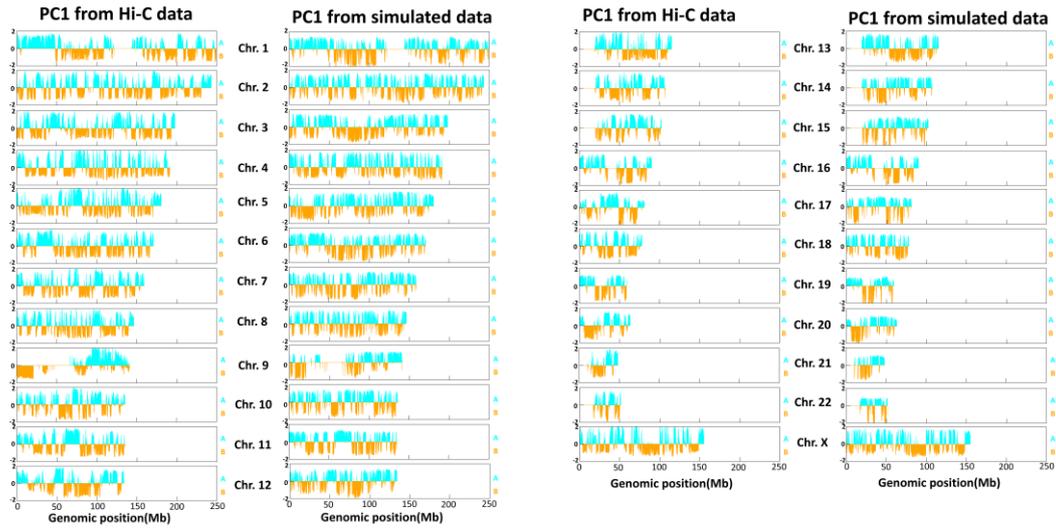

**Figure S3. Chromosome-wise PC1 profiles from simulated and experimental Hi-C data in K562 cells.** Principal component 1 (PC1) values were computed for each chromosome from correlation matrices of simulated and experimental Hi-C contact maps to delineate A/B compartments. The profiles include all autosomes and the X chromosome. These data correspond to the Pearson correlation analysis presented in Fig. 2D and 2E.

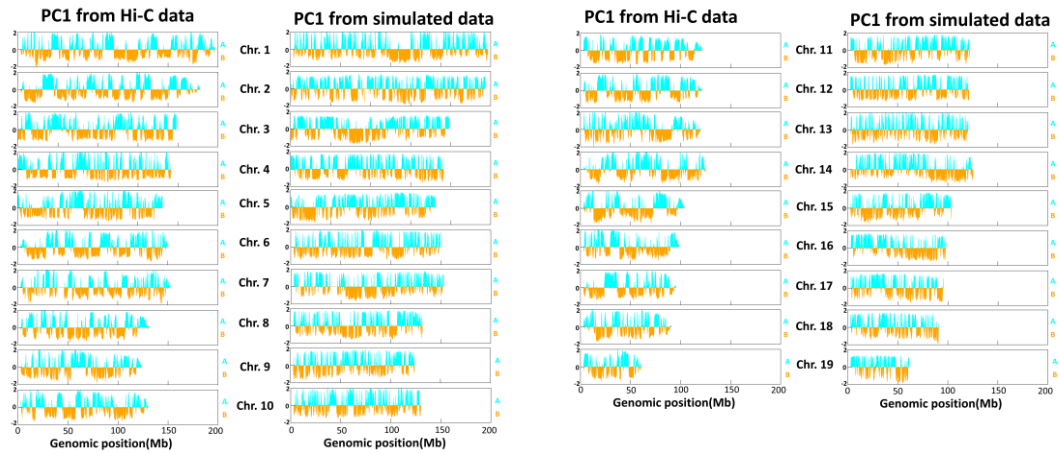

**Figure S4. Chromosome-wise PC1 profiles from simulated and experimental Hi-C data in the 8-cell stage of mouse development.** Principal component 1 (PC1) values were computed for each chromosome from correlation matrices of simulated and experimental Hi-C contact maps to delineate A/B compartments. The profiles include all autosomes. These data correspond to the Pearson correlation analysis presented in Figure S1D and S1E.

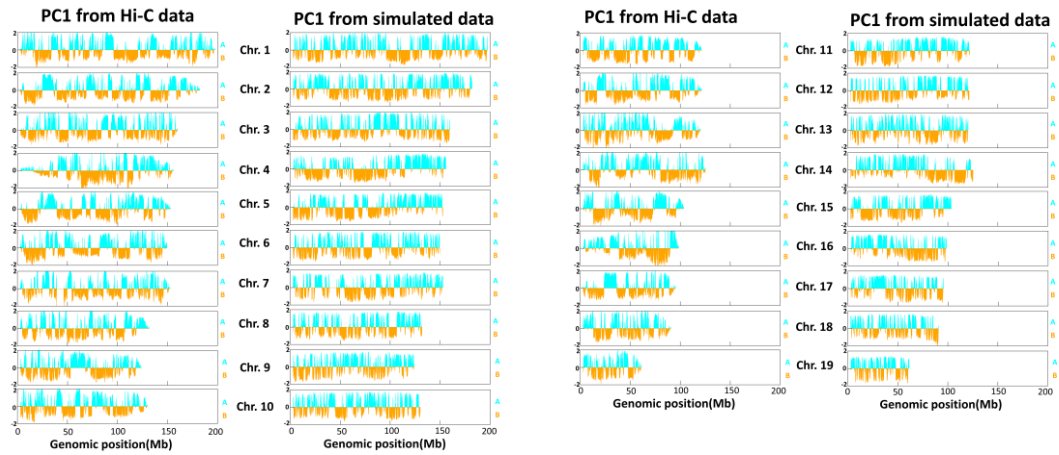

**Figure S5. Chromosome-wise PC1 profiles from simulated and experimental Hi-C data in the inner cell mass (ICM) stage of mouse development.** Principal component 1 (PC1) values were computed for each chromosome from correlation matrices of simulated and experimental Hi-C contact maps to delineate A/B compartments. The profiles include all autosomes. These data correspond to the Pearson correlation analysis presented in Figure S2D and S2E.

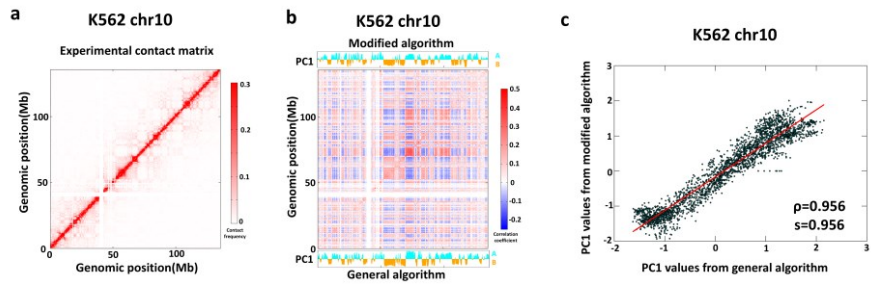

**Figure S6. Validation of the method for identifying A/B compartments from sparse Hi-C contact data. (A)** Experimental Hi-C contact matrix of K562 chromosome 10. **(B)** Principal component 1 (PC1) values calculated using the modified algorithm (top) and the general algorithm described in Section VI (bottom), both based on the contact matrix in (A). **(C)** Pearson correlation analysis between the two PC1 profiles shown in (B), demonstrating strong concordance ($\rho$=0.956).

**Captions of Movies:**

Movie 1: Rotated view of a representative initial structure used in the simulation of artificial chromatin fiber in both dynamic and static models, featuring two types of blocks depicted in distinct colors

Movie 2: Rotated view of a representative final structure from the dynamic model simulation.

Movie 3: Evolution of the artificial chromatin structure along a representative folding trajectory in the dynamic model.

Movie 4: Rotated view of a representative final structure from the static model simulation.

Movie 5: Evolution of the artificial chromatin structure along a representative folding trajectory in the static model.

Movie 6: Evolution of the artificial chromatin structure along a representative folding trajectory in the static model, starting from a structure with segregated red and blue blocks.